# Enhancement of Transition Temperature in $Fe_xSe_{0.5}Te_{0.5}$ Film via Iron Vacancies

J. C. Zhuang[1,2], W. K. Yeoh[2,3,4,a)], X. Y. Cui[3,4], J. H. Kim[2], D. Q. Shi[2], Z. X. Shi[1,a)], S. P. Ringer,[3,4] X. L. Wang[2], S. X. Dou[2]

[1]*Department of Physics and Key Laboratory of MEMS of the Ministry of Education, Southeast University, Nanjing 211189, People's Republic of China*

[2]*Institute for Superconducting and Electronic Materials, University of Wollongong, North Wollongong, New South Wales 2500, Australia*

[3]*Australian Centre for Microscopy and Microanalysis, University of Sydney, Sydney, New South Wales 2006, Australia*

[4]*School of Aerospace, Mechanical and Mechatronic Engineering, University of Sydney, New South Wales 2006, Australia*

**Abstract**

The effects of iron deficiency in $Fe_xSe_{0.5}Te_{0.5}$ thin films ($0.8 \leq x \leq 1$) on superconductivity and electronic properties have been studied. A significant enhancement of the superconducting transition temperature ($T_C$) up to 21 K was observed in the most Fe deficient film ($x = 0.8$). Based on the observed and simulated structural variation results, there is a high possibility that Fe vacancies can be formed in the $Fe_xSe_{0.5}Te_{0.5}$ films. The enhancement of $T_C$ shows a strong relationship with the lattice strain effect induced by Fe vacancies. Importantly, the presence of Fe vacancies alters the charge carrier population by introducing electron charge carriers, with the Fe deficient film showing more metallic behavior than the defect-free film. Our study provides a means to enhance the superconductivity and tune the charge carriers via Fe vacancy, with no reliance on chemical doping.

a) Authors to whom correspondence should be addressed. Electronic addresses: waikong.yeoh@sydney.edu.au and zxshi@seu.edu.cn.

Shortly after the discovery of superconductivity in LaFeAsO$_{1-x}$F$_x$ with a critical transition temperature, $T_C$, of 26 K,[1] superconductivity was observed in PbO-type FeSe.[2] Due to its simple crystal structure, (composed of a stack of superconducting Fe$_2$Se$_2$ layers along the $c$-axis), the FeSe system has attracted tremendous interest for exploring the mechanism of high temperature superconductivity. Even though the $T_C$ of FeSe is as low as 8 K, it can be substantially improved either by chemical doping or by the application of pressure. For instance, partial substitution of Te for Se (chemical pressure effect) leads to an increase in $T_C$ up to ~ 15 K with $0.3 \leq x \leq 0.7$ for FeSe$_{1-x}$Te$_x$ compounds,[3,4] while application of external pressure of 8.9 GPa (Refs. 5,6) leads to enhancement of $T_C$ up to 36.7 K for pure FeSe samples. The other $T_C$ enhancement option is via the strain effect, where a maximum $T_C$ of 21 K can be obtained in FeSe$_{0.5}$Te$_{0.5}$ thin film through lattice mismatch with different substrates.[7] More recently, a record high $T_C$ of 65 K has been reported for monolayer FeSe film by two separate groups.[8,9] These studies suggest that the charge carrier population and structural variation induced by strain could be the two major parameters that are critical to the high $T_C$ of the Fe based superconductors.[10] Since excess Fe is generally deleterious to superconductivity,[11] low Fe content samples ($x < 1$, Fe$_x$Se$_{0.5}$Te$_{0.5}$), which theoretically should induce hole carriers by the simple electron counting rule and chemical pressure, provide the ideal opportunity to verify the roles of these factors. Nevertheless, little research on samples with $x < 1$ has been reported. In fact, the work on Fe deficiency is rather contradictory. Both Sudesh $et\ al.$[12] and Sala $et\ al.$[13] reported no correlation of $T_C$ with reduction of iron content. While Bendele $et\ al.$[14] showed $T_C$ enhancement in the $x < 1$ region, Chen $et\ al.$[15] found that Fe$_4$Se$_5$ ($x = 0.8$) is a non-superconducting phase. Evidently, the influence of Fe vacancy on superconductivity is still unclear. This has motivated us to carry out a systematic study of the effects of Fe deficiency on superconductivity and structural evolution.

In this work, we investigate the variation of structural properties and superconducting properties, as well as variation in the population of charge carriers, among film samples with different Fe content. Our results show that lattice strain is induced by Fe vacancy in the low Fe content region. A $T_C$ around 21 K is obtained in Fe$_{0.8}$Se$_{0.5}$Te$_{0.5}$ film, which is related to lattice strain.

Three kinds of polycrystalline pellets of $Fe_xSe_{0.5}Te_{0.5}$ with nominal compositions of $FeSe_{0.5}Te_{0.5}$ ($x = 1$), $Fe_{0.9}Se_{0.5}Te_{0.5}$ ($x = 0.9$), and $Fe_{0.8}Se_{0.5}Te_{0.5}$ ($x = 0.8$) were fabricated as targets. Powders of Fe, Se, and Te were mixed together in the stoichiometric ratios and heated in an evacuated quartz tube at 850 ºC for 12 h. After the sintering, the mixture was reground, pelletized, and sintered in the evacuated quartz tube at 400 ºC for 6 h to make the target dense. The films were grown under vacuum conditions (~$6 \times 10^{-4}$ Pa) by pulsed laser deposition (PLD) using a Nd: YAG laser (wavelength: 355 nm, repetition rate: 10 Hz). Single crystal $CaF_2$ (100) with lattice parameter $a_0 = 5.463$ Å was selected as the substrate due to its non-oxide nature and the low mismatch between its lattice parameter ($a_0/\sqrt{2} = 3.863$ Å) and the $a$-axis parameter of the film (around 3.8 Å). The deposition temperature was set at 450 ºC, and the laser energy was 200 mJ/pulse. The substrate-target distance was maintained at 4 cm, and the deposition time was the same for all three samples. The thickness of the films, measured by scanning electron microscopy, was around 50 nm. The crystal structure and orientation of the films were characterized by X-ray diffraction (XRD) at room temperature. Electrical resistivity ($\rho$) and Hall measurements were carried out on a 14 T physical properties measurement system.

The inset of Fig. 1(a) displays the powder XRD results for the three kinds of targets, normalized by the value of the intensity of the respective (101) peaks for direct comparison. All the main peaks in the three bulk samples can be well indexed based on the PbO tetragonal structure. While the $x = 1$ sample shows no trace of impurity, chalcogen ($Ch$) rich impurity phases of $FeTe_2$ and $Fe_7Se_8$ are observed in the samples with $x = 0.9$ and 0.8. As there is no shift of the peak position of the tetragonal phase for any of the three targets, and the intensity of the impurity phase peaks increases with decreasing Fe content, it is suggested that the Fe deficiency in the bulk is compensated by the formation of the impurity phases. The presence of impurity is consistent with the reported phase diagram of FeSe compounds, where $Fe_7Se_8$ coexists with PbO-type FeSe in the $x < 1$ region.[11] Fig. 1(a) shows the XRD results for thin films deposited from different targets normalized by the intensity of the respective (001) peaks. These films are denoted by the nominal composition of the target. Only the (00$l$) reflections of the films and of the substrate are present, indicating the out-of-plane orientation of the structure in these films. Fig. 1(b) shows the XRD pattern of the bare substrate in order to separate the substrate peaks and film peaks. The impurity phases that exist in the polycrystalline target are no longer observed in the XRD patterns of the films. Interestingly, a significant shift in the peak positions is displayed, showing that the structural lattice parameter $c$ shrinks from 5.98 Å for $x = 1$ to 5.80 Å for $x = 0.8$, which is smaller than for the bulk polycrystalline targets (6.01 Å) or FeSe$_{0.5}$Te$_{0.5}$ single crystal (6.05 Å),[16] but in agreement with the previous reports on films.[7,17] The normalized intensities of the (00$l$) reflections (except for the (001) peak) become weaker and are broadened towards higher scanning angles with decreasing Fe content, which may be due to Fe disorder or inhomogeneity.[18] The lattice mismatch between the CaF$_2$ substrate ($a_0/\sqrt{2}$ = 3.863 Å) and FeSe$_{0.5}$Te$_{0.5}$ ($a$ = 3.8 Å) is as small as 1.66%, much smaller than for other substrates (e.g. MgO 10.84% and SrTiO$_3$ (STO) 2.76%). Considering that in the larger lattice mismatched films, there is no clear correlation between the lattice mismatch and the lattice parameters of the film,[19,20] and with all the films being prepared on same CaF$_2$ substrate in this paper, the lattice mismatch effect on the lattice parameters will be insignificant. In fact, we believe that the observed variation of the out-of-plane orientation is dominated by the composition of the films rather than the tensile strain induced by the lattice mismatch. Furthermore, based on density-functional theory (DFT) calculations, there are two types of point defects with similar low formation energy under the Fe-deficient conditions: Fe vacancy and Se/Te interstitial. Nevertheless, these two phases show opposite structural transitions. Fe vacancy phase leads to a smaller $c$ lattice parameter (by 0.036 Å), while Se/Te interstitial phase leads to a larger $c$ (by 0.25 Å/0.60 Å) in comparison with the stoichiometric phase ($x = 1$). Combining the XRD results and the DFT results, the Fe vacancy scenario is the more favored defect compared to the Se/Te interstitial in our low Fe content samples.

Fig. 2(a) shows the electronic resistivity versus temperature ($R$-$T$) curves from 300 K to 10 K for the three films. From the $R$-$T$ curves, the film with the least Fe ($x = 0.8$) has the lowest normal state resistivity, confirming that resistivity drops with decreasing Fe content. A hump can be observed in the normal state resistivity. It shifts to a higher temperature with decreasing Fe content, making the $Fe_{0.8}Se_{0.5}Te_{0.5}$ sample display a more metallic behavior than the $Fe_{1.0}Se_{0.5}Te_{0.5}$ sample. In order to provide a clearer picture on the shift of the hump position, $\partial(R/R_{300\,K})/\partial(T)$ curves are shown in the inset of Fig. 2(b). If we use the criterion of the point where $\partial(R/R_{300\,K})/\partial(T) = 0$ as the hump position, the temperature of the hump shifts from ~ 200 K for $x = 1$ to ~275 K for $x = 0.8$. $\partial(R/R_{300\,K})/\partial(T) = 0$ corresponds to the transition from semiconducting to metallic-like behavior, based on the normal state resistivity. As such, the hump position can be an indication of electronic property transition. The inset of Fig. 2(a) displays an enlarged view of the $R$-$T$ curves in the temperature range from 14 K to 24 K. The $T_C^{onset}$ is determined by 90 % of the normal state resistivity for all three samples. One of the most interesting features in the $R$-$T$ curve is the significant enhancement of $T_C$ from ~16 K for $x = 1$ to ~21 K for $x = 0.8$. The $T_C$ enhancement is also accompanied by broadening of the transition temperature and lower peak intensity, indicating the presence of multiple superconducting phases. According to Chen et al.[15] and Bendele et al.[14], higher $T_C$ can be obtained in lower Fe content samples compared with $Fe_{1.01}Se$, which is consistent with our results. In particular, it has been reported that at least three orders of Fe-vacancy can be found in the FeSe sample: $\beta$-$Fe_3Se_4$ ($x = 0.75$), $\beta$-$Fe_4Se_5$ ($x = 0.80$), and $\beta$-$Fe_9Se_{10}$ ($x = 0.90$).[15] Combining the XRD and DFT results, along with the interpretation of Ref. 15, the observed broadening of the superconducting transition in the current work is most likely due to the Fe disorder effect induced by the inhomogeneous distribution of Fe vacancy in the $x < 1$ region.

In order to understand how the charge carrier population varies with the Fe content, we performed Hall measurements. The inset of Fig. 3 shows the transverse resistivity, $\rho_{xy}$, at different temperatures for $x = 0.9$. An anomalous Hall effect (AHE), which has been reported by Feng et al.[21,22] and Tsukada et al.[23] in FeSe films, was observed by sweeping the magnetic field from -1 T to 1 T. Although the origin of the AHE remains elusive, the observed AHE can be attributed to the presence of spontaneous magnetization in FeSe thin films.[21, 22] Nevertheless, the Hall coefficient, $R_H = \rho_{xy}/B$, where $B$ is the magnetic flux density, as determined by linear fitting of the $\rho_{xy}$ curves between 1 T $\leq$ $|B|$ $\leq$ 4 T, is in a magnetic field range free of AHE influence. Fig. 3 shows that the $R_H$ of all three samples is almost temperature independent from room temperature down to 80 K and has a value around $2 \times 10^{-9}$ m$^3$/C. This phenomenon is similar to what has been previously reported.[24] The sign of $R_H$ changes from positive for $x = 1$ and 0.9 to negative for $x = 0.8$ in the low temperature region. Such a sign reversal was claimed as evidence for the multiband nature of the band structure in the 11 system.[24] Both the DFT simulations and angle-resolved photoemission spectroscopy reveal one inner closed Fermi pocket and two outer cylindrical Fermi surfaces near $\Gamma$ (hole sections), and two electron-like Fermi surfaces near the $M$ point.[25,26] The classical formula for the Hall coefficient of semiconductors in the presence of both electron- and hole-type carriers is used:[27]

$$R_H = \frac{1}{e} \frac{(\mu_h^2 n_h - \mu_e^2 n_e) + (\mu_h \mu_e)^2 B^2 (n_h - n_e)}{(\mu_e n_h + \mu_h n_e)^2 + (\mu_h \mu_e)^2 B^2 (n_h - n_e)^2} \quad (1)$$

where $\mu_h$, $\mu_e$, $n_h$, and $n_e$ are hole mobility, electron mobility, hole density, and electron density, respectively. Eq. (1) predicts that $R_H = e^{-1}(\mu_h^2 n_h - \mu_e^2 n_e)/(\mu_e n_h + \mu_h n_e)^2$ when $B \to 0$, and $R_H = e^{-1} 1/(n_h - n_e)$ in the limit of $B = \infty$. It can be obtained from the sign reversal of $R_H$ that the population of electron carriers increased in the Fe vacancy samples. This is in conflict with the general assumption from the electron counting rule that Fe vacancy introduces hole carriers. Although no chemical doping was involved, reducing the Fe content can be considered as "self-doping", as it involves changing the electronic properties, as shown in the resistivity results. This is similar to the Fe vacancy disorder in the K$_x$Fe$_{2-y}$Se$_2$ case, where the disorder effect raises the chemical potential significantly, giving rise to enlarged electron pockets similar to those in a highly doped system, but without adding carriers to the system.[28]

While more research is needed to explore the relation between the charge carriers and high $T_C$ in Fe vacancy samples, our results, however, including the decreased $c$ lattice parameter, the enhancement of $T_C$, and the shift of the hump position in the $R$-$T$ curves, consistently show that Fe vacancy has similar effects to high-pressure on the structural and superconducting properties in the 11 system.[29-31] As a result, it is likely that the origin of the enhancement of $T_C$ is strongly correlated with the lattice strain induced by the Fe vacancy disorder. In fact, the effects of lattice strain on $T_C$ have long been observed in polycrystalline and thin film superconductors.[32-34] Lattice strain evoked by Mg vacancies in the MgB$_2$ system decreases the $T_C$ by around 2 K.[32] Moreover, superconductivity was induced in the non-superconducting parent compounds of BaFe$_2$As$_2$ and FeTe by controlling the interfacial tensile lattice strain between the superconducting film and the substrate or buffer layer.[33,34] In general, there are two types of lattice strain, uniform and non-uniform strain, that can be present in a crystal.[35] Uniform strain causes the unit cell to behave in an isotropic way, accompanied by the shifting of peaks in XRD, while non-uniform strain, which can be stimulated by point defects, plastic deformation, or poor crystallinity, leads to peak broadening. In the current work, both kinds of strain are found to exist in the Fe vacancy films, based on the XRD results, and show a strong correlation with the $T_C$. For instance, the sample with the smallest lattice parameter (high uniform strain) displays the highest $T_C$. On the other hand, peak broadening (non-uniform strain) becomes more prominent in Fe vacancy samples. The non-uniform strain values have been estimated from the slope of the Williamson-Hall (WH) plot of the calculated full width at half maximum ($FWHM$) $\times \cos\theta$ as a function of $\sin\theta$, where $\theta$ is the Bragg angle.[36] It can be seen in Fig. 4(a) that all three curves yield a straight line, which is characteristic of Lorentzian profiles.[36,37] Fig. 4(b) shows the $T_C$ dependence on both uniform and non-uniform strain. A strong correlation between $T_C$ and lattice strain is observed, where high $T_C$ values are located in the high strain regions. The current results combined with all these studies [32-34] lead to the conclusion that lattice strain plays an important role in the exploration of superconductivity or $T_C$ enhancement.

In summary, we have investigated the structural and superconducting properties of Fe-deficient Fe$_x$Se$_{0.5}$Te$_{0.5}$ thin films ($x$ down to 0.8). Our work provides two critical discoveries on the 11 system: i) Fe vacancy film can be fabricated under controlled conditions, and ii) samples with lower Fe content possess higher $T_C$. With the recent discovery suggesting that the Fe$_{1-x}$Se phases with iron vacancy phases are more likely to be the parent phase of the 11 system,[15] instead of the previously suggested Fe$_{1+\delta}$Te, this work will further illuminate the phase diagram and the role of Fe in the superconductivity.

**Acknowledgements**

This work was supported by the Natural Science Foundation, the Ministry of Science and Technology of China (973 project: No. 2011CBA00105), Scientific Research Foundation of Graduate School (Grant No. YBJJ1314) of Southeast University, and the Australian Research Council through Discovery Project DP 120100095.

Figure captions

FIG. 1. (a) XRD patterns of $Fe_xSe_{0.5}Te_{0.5}$ thin films with $x$ = 1, 0.9, 0.8 deposited on $CaF_2$ (100) substrate. "♦" marks the reflections of impurity phases in the substrate, identified by ICDD card. (b) XRD reflection of bare substrate. Inset: XRD results for $Fe_xSe_{0.5}Te_{0.5}$ targets.

FIG. 2. (a) Electronic resistivity versus temperature (*R-T*) curves from 300 K to 10 K for the three films. Inset: Enlarged view of *R-T* curves from 14 K to 24 K. (b) $\partial(R/R_{300\,K})/\partial(T)$ curves for the three films from 10-50 K. Inset: $\partial(R/R_{300\,K})/\partial(T)$ curves for the temperature range from 50 K to 300 K. Arrows point to the hump position, defined as the point where $\partial(R/R_{300\,K})/\partial(T) = 0$.

FIG. 3. Temperature dependence of the Hall coefficient for the three films. Inset: linear relationship between $\rho_{xy}$ and $B$ at different temperatures.

FIG. 4. (a) Williamson-Hall plots: $FWHM \times \cos\theta$ as a function of $\sin\theta$. (b) $T_C$ dependences of both uniform strain (using the parameter 1-$c/c_0$, where $c_0$ is the *c*-axis lattice parameter of $FeSe_{0.5}Te_{0.5}$ film) and non-uniform strain from the WH plots. The dashed line is only to guide the eye.

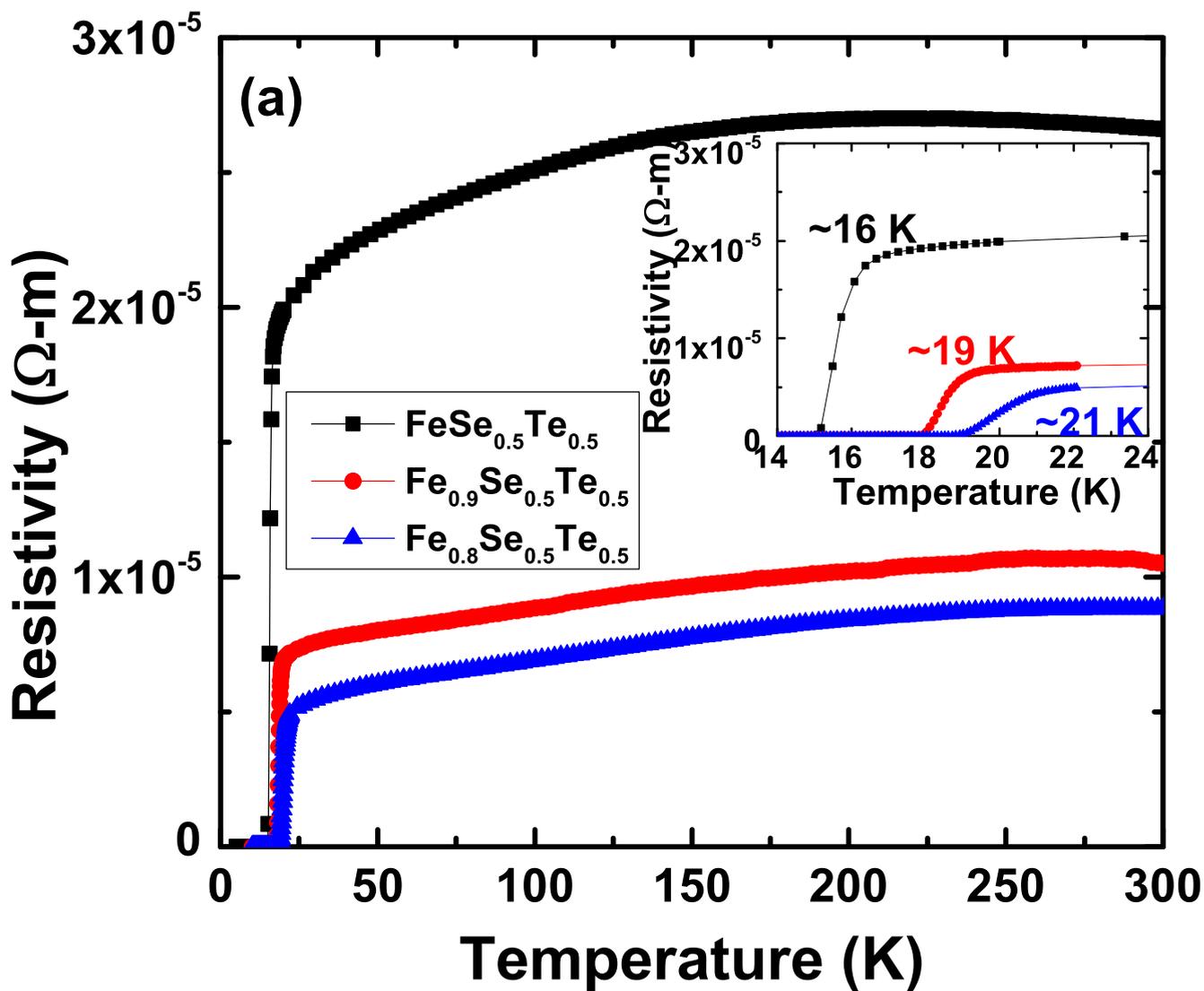

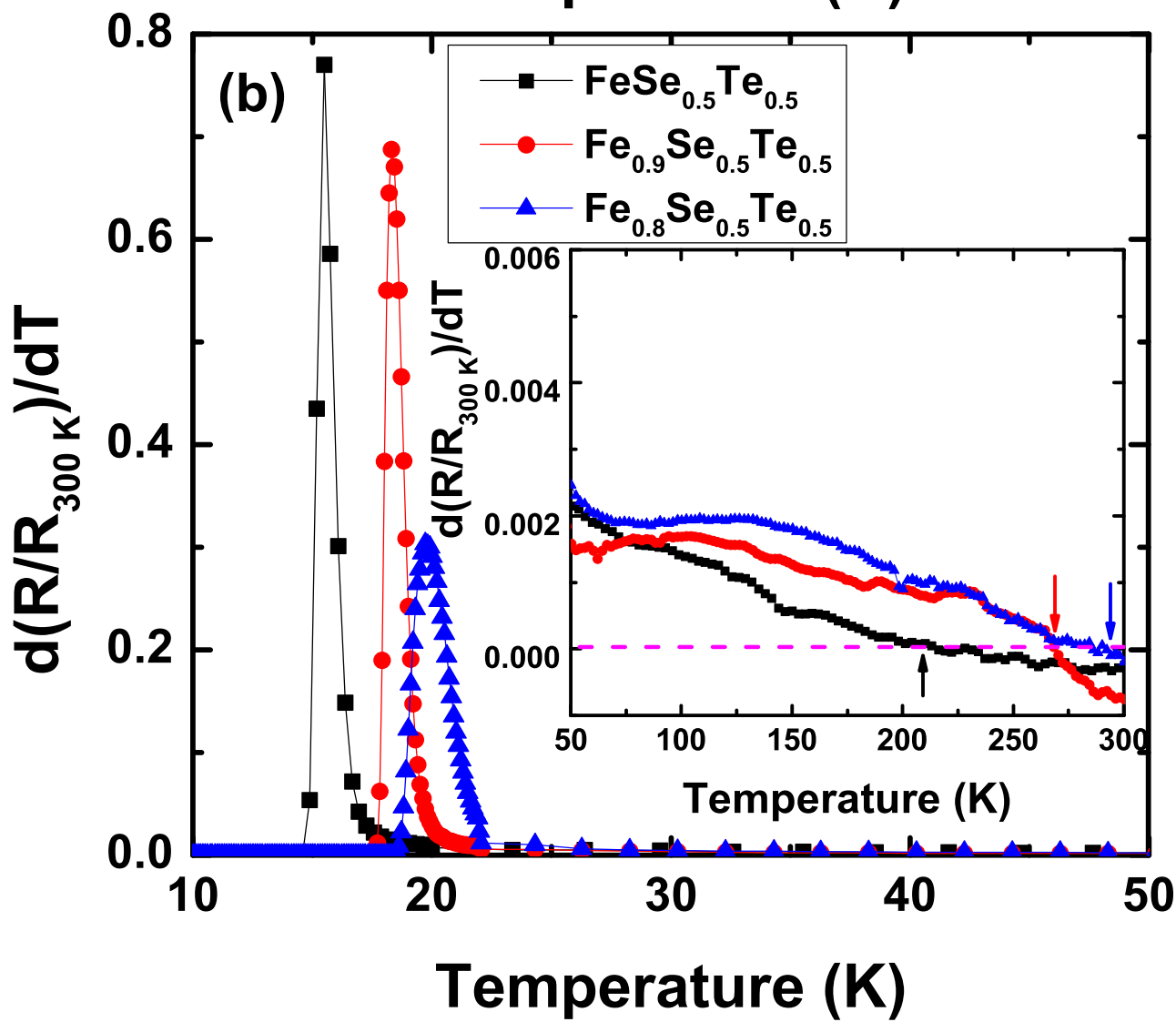

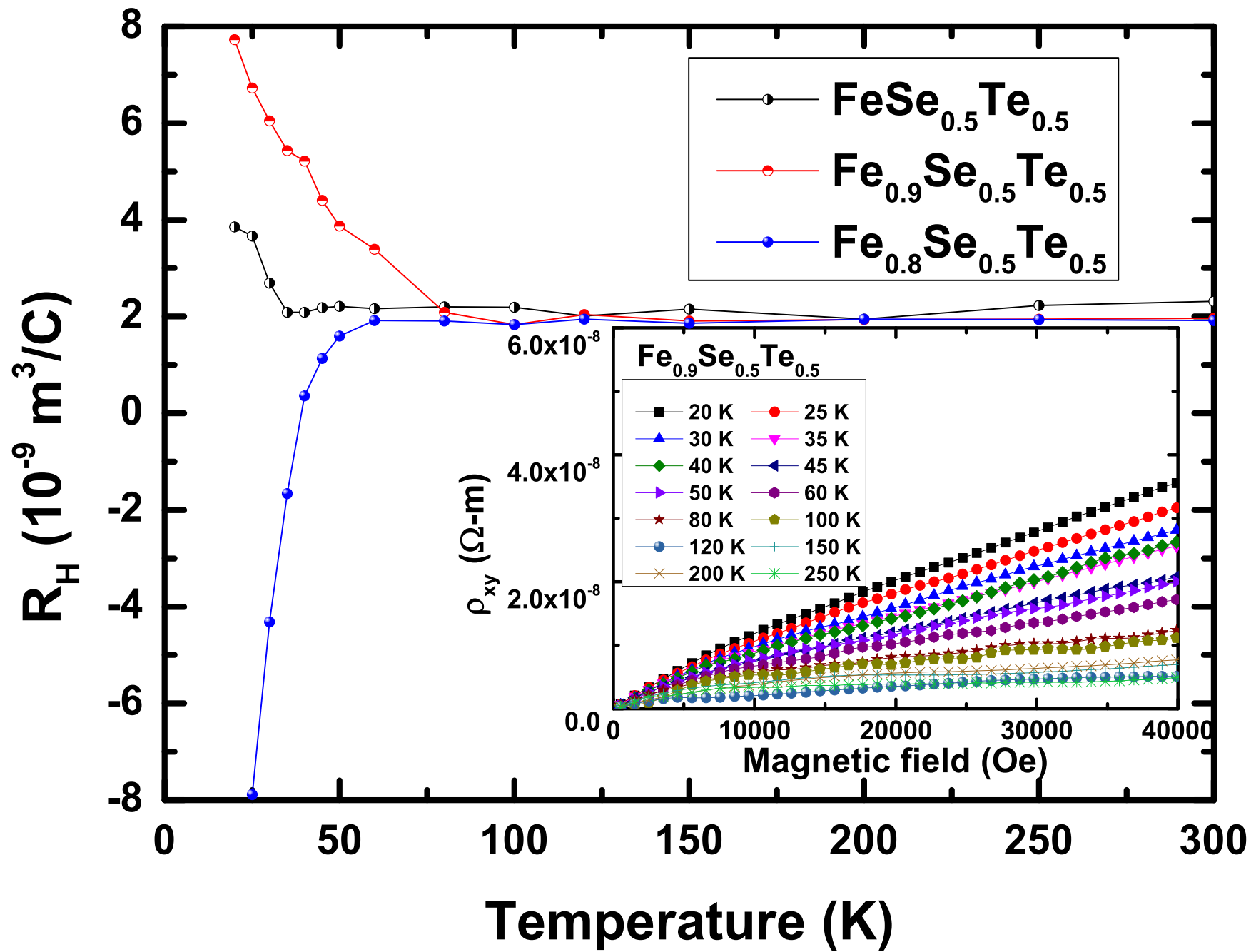

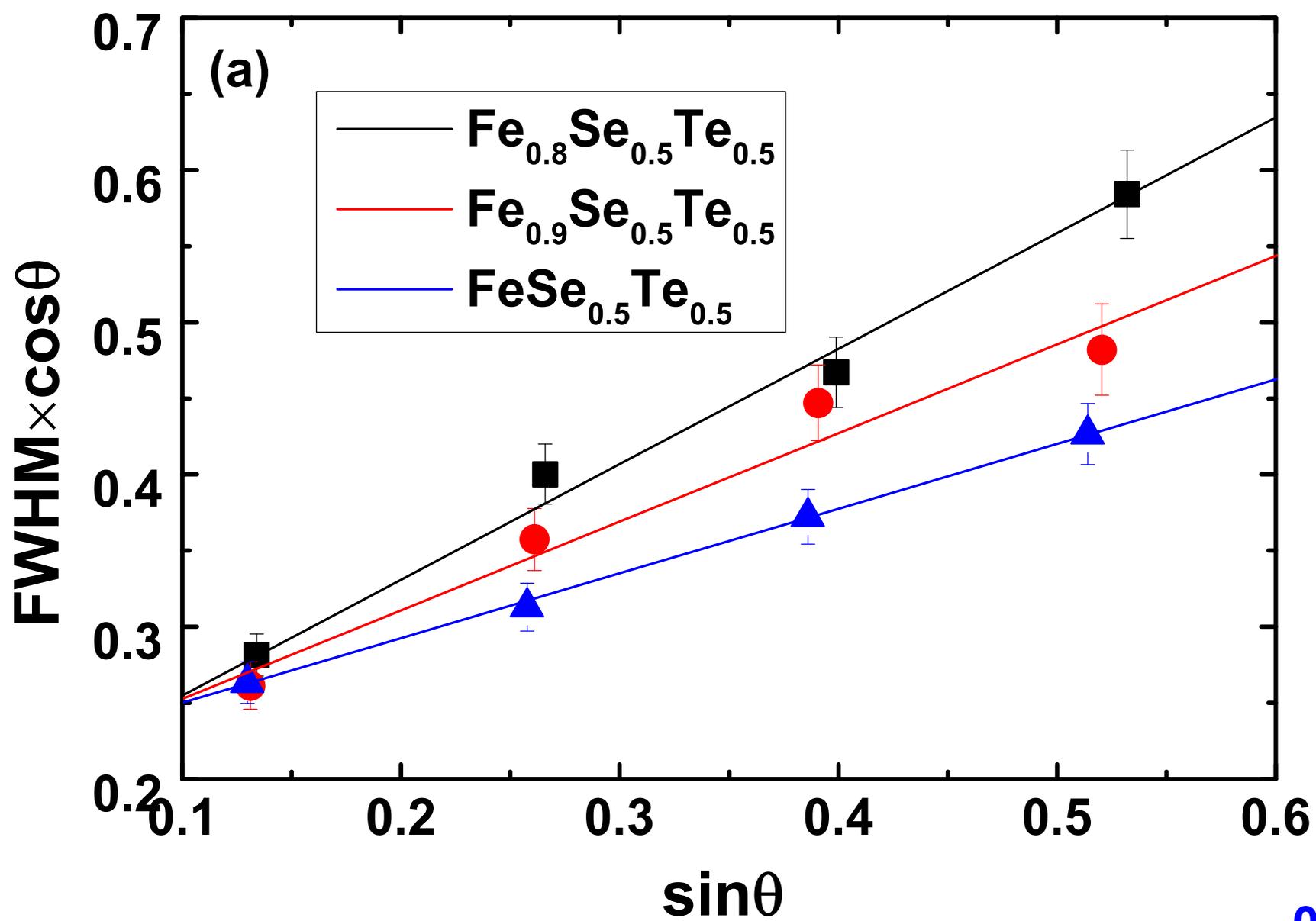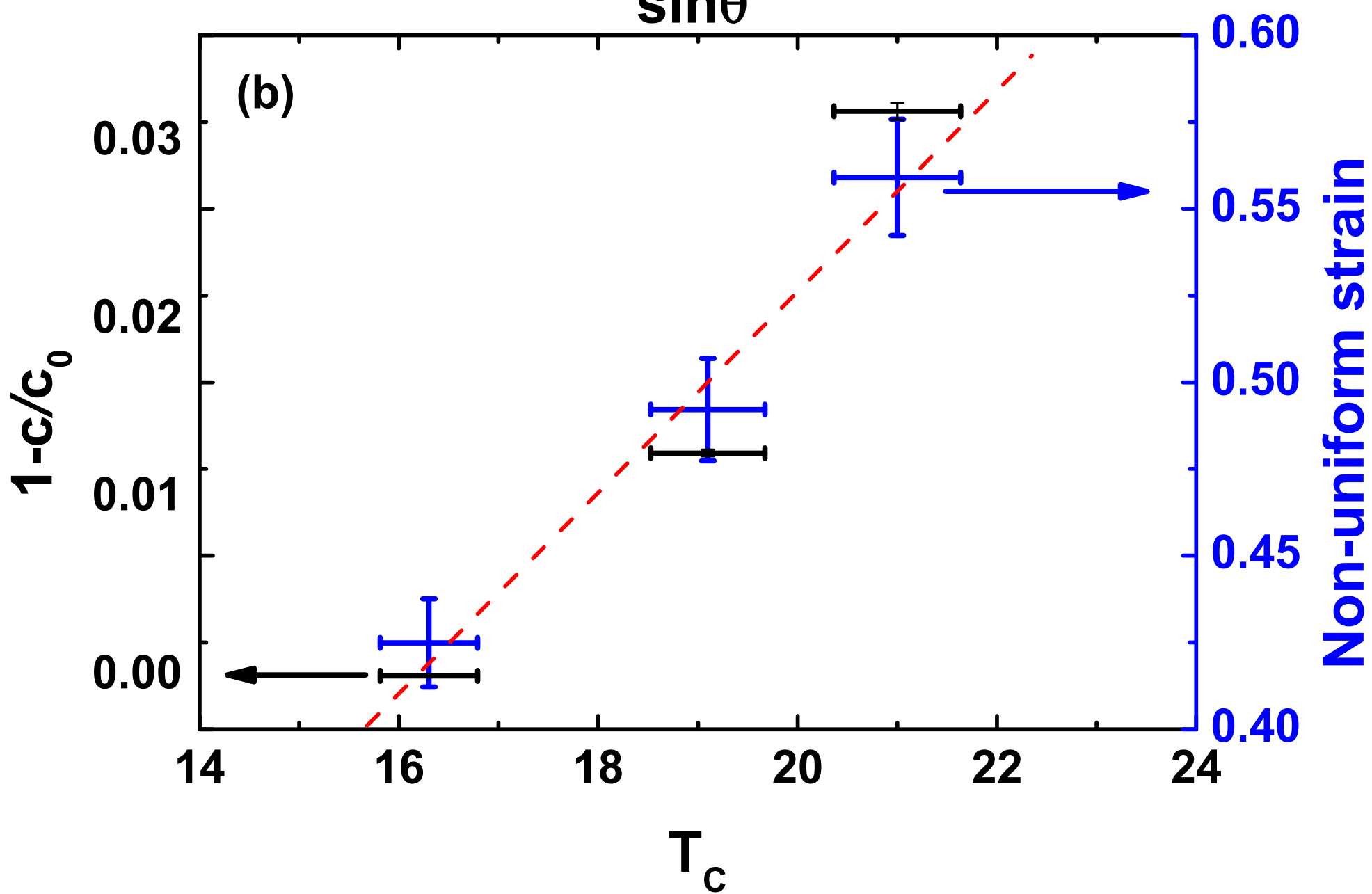